\documentclass[12pt]{article}

\begin{document}

\title{Which Physics Laws are Deduced from the Logic Properties of the 
Information?}
\author{G.Quznetsov \\
quznets@geocities.com}
\maketitle

\begin{abstract}
The relativity theory principles and the quants theory principles are
deduced from logic properties of the information, obtained from a physics
device.
\end{abstract}

This paper presents a logic development of the Bergson \cite{A1}, Whitehead 
\cite{A2}, Capek \cite{A3}\cite{A4}, Stapp \cite{A5}-\cite{A8}, Whipple \cite
{A9} ideas on ''... events must be treat as the fundamental objective
constituents ... events and not particles constituite the true objective
reality''\cite{A10}. (The A.Jadczyk and Ph.Blanchard papers \cite{A11}-\cite
{A13} are related to this topic for some time past.):

An information, which is obtained from a physics device $\widehat{\mathbf{a}}
$, can be expressed by a set $\mathbf{a}$ of any language sentences. The set 
$\mathbf{a}$ is denoted as ''the recorder of the device $\widehat{\mathbf{a}}
$''. The set of the recorders call into existence structures, similar to a
clocks . The following results are deduced from the logic properties of the
recorders set \cite{A15}:

First, all such clocks have got the same direction, i.e. if the event,
expressed by the sentence $A$, precedes to the event, expressed by the
sentence $B$, with respect to any such clock, then it is the same for all
other such clocks.

Second, the Time, defined by such clocks, proves irreversible, i.e. no the
recorder can obtain the information, that a certain event has taken place,
before it has actually taken place. Thus, nobody can return back into the
Past Times or obtain the information from the Future Times.

Third, the set of recorders has been embedded in the metric space by some
natural method; i.e. all metric space axioms are obtained from the logic
properties of the recorder set.

Fourth, if this metric space proves to be the Euclidean space, then the
corresponding recorders ''space-time'' obeys the Poincare complete group
transformations. I.e. in this case the Special Theory Relativity follows
from the logic properties of the information. If this metric space is not
Euclidean, then any non-linear geometry exists on the space of the
recorders, and any variant of the General Relativity Theory can be realized
on this space.

Therefore, the principal time properties - the one-dimensionality and the
irreversibility -, the space metric properties and the spatial-temporal
principles of the theory of the relativity are deduced from the logic
properties of the recorders set. Hence, if you have got any set of the
objects, which able to get, to keep and/or to give any information, then
''the time'' and ''the space'' are inevitable on this set. And it is all the
same: or this set is in our world or this set is in any other worlds, in
which the spatial- temporal structure does not exist initially. Hence, the
spatial-temporal structure arises from the logic properties of the
information.

There is the evident nigh affinity between the classical probability
function and the Boolean function of the classical propositional logic \cite
{A16}. These functions are differed by the range of value, only. That is if
the range of values of the Boolean function shall be expanded from the
two-elements set $\left\{ 0;1\right\} $ to the segment $[0;1]$ of the real
numeric axis then the logic analog of the Bernoulli Large Number Law can be
deduced from the logic axioms. And if the range of values of such function
shall be expanded to the segment of some suitable variant of the hyperreal
numeric axis then this theorem shall insert some statistical meaning for
this function \cite{A17}.

The probability must comply with certain simple condition in order to be
expressed by a relativistic $\mu +1$-vector of the probability density \cite
{A18}. Such probability is denoted as ''the trackelike probability''. The
Dirac equation is deduced from such probability properties by the Poincare
group transformations \cite{A19} \cite{A20}. Hence the physics elementary
particle behavior in the vacuum looks like to the trackelike probability
behavior. In the two- slits
experiment  if the partition with two slits between the source of the
physics particle and the detecting screen exists in the vacuum then the
interference of the probability is observed. But if this system shall be
placed in the Wilson cloud chamber then the particle shall got the clear
trace, marked by the condensate drops, and whole interference shall vanished.
It looks like to the following: the physics particle exists in the moment,
only, in which some event on this particle is happening. And in other times
this particle does not exist and the probability of some event on this
particle exists, only. 

Hence, if an events on this particle do not happen
between the event-birth and the event-detection then the particle behavior
is the probability behavior between these events, and the interference is
visible. But in the Wilson cloud chamber, where the ionization acts form
the almost continuous line, the particle has got the clear trace and no the
interference. And the particle moves because such line is not absolutely
continuous. Every point of the ionization act has got the neighboring
ionization point, and the event on this particle is not happen between these
points. Therefore, the physics particle moves because the corresponding
probability is propagated in the space between these points.

Therefore a particle is an ensemble of events, bounded by a probabilities 
(that is similar to \cite{A21}).

In the $3+1$ space-time all interactions between fermions can be expressed
by some division algebra (the Cayley algebra) but such algebra does not exist 
in thespace- time with more than $3+1$ dimension \cite{A22}. Hence the fermions
can not go out from this $3+1$ space-time.

Thus particles and fields are not the basic entities of Universe but the
logic events and the logic probabilities are the basic entities. Universe -
i.e. the time, the space and whole their contents - is the by-product of the
deduction from the logic events.

\end{document}